\documentclass[12pt]{article}
\usepackage{graphicx}
\usepackage{lineno}
\newcommand\pubdate{\today}

\def\Title#1{\begin{center} {\Large #1 } \end{center}}
\def\Author#1{\begin{center}{ \sc #1} \end{center}}
\def\Address#1{\begin{center}{ \it #1} \end{center}}

\newcommand\pubblock{\rightline{\begin{tabular}{l}  \\ % Author's note number [if you need to add one] goes here
         \pubdate  \end{tabular}}}
\newenvironment{Abstract}{\begin{quotation}  }{\end{quotation}}
\newenvironment{Presented}{\begin{quotation} \begin{center} 
             PRESENTED AT\end{center}\bigskip 
      \begin{center}\begin{large}}{\end{large}\end{center} \end{quotation}}
\begin{document}

%%\linenumbers

\begin{titlepage}
 \pubblock
\vfill
\Title{Searching for additional Higgs bosons at ATLAS}
\vfill
\Author{Anna Kaczmarska \\
         on behalf of the ATLAS Collaboration}
\Address{Institute of Nuclear Physics PAN, Cracow, Poland}
\vfill
\begin{Abstract}
Extending the Higgs sector by introducing additional scalar fields to account for the electroweak symmetry breaking, can provide solutions to
some of the questions the Standard Model fails to answer.
Introducing additional scalar fields leads to extra Higgs-like particles, which can be either neutral or charged.
These proceedings present some recent direct searches for additional Higgs bosons, using proton–proton collision data at $13$~TeV collected
by the ATLAS experiment in Run 2 of the LHC.
\end{Abstract}
\vfill
\begin{Presented}
DIS2023: XXX International Workshop on Deep-Inelastic Scattering and
Related Subjects, \\
Michigan State University, USA, 27-31 March 2023 \\
     \includegraphics[width=9cm]{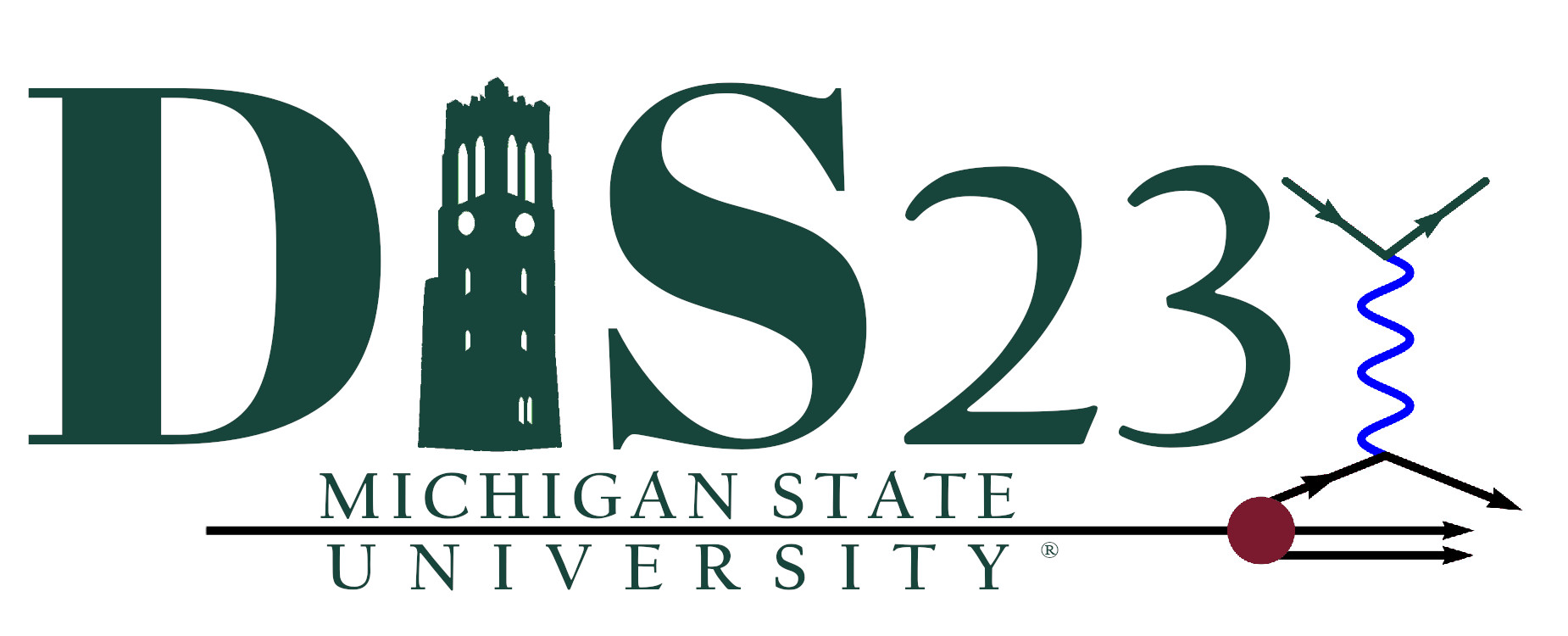}
\end{Presented}
\vfill
\small{Copyright [2023] CERN for the benefit of the ATLAS Collaboration. CC-BY-4.0 license.}
\end{titlepage}

%%%%%%%%%%%%%%%%%%%%%%%%%%%%%%%%%%%%%%%%%%%%%%%%%%%%%%%%%%%%%%%%%%
\section{Introduction}

Since the discovery of the Higgs boson by the CMS and ATLAS Collaborations in 2012, its properties have been measured to increasing precision. 
So far, an excellent agreement with the predictions for a Standard Model (SM) Higgs boson is observed.
However, the SM, while highly successful, is not considered to be a complete theory as it is not capable of explaining 
some of the phenomena seen in nature.
Extending the Higgs sector by introducing additional scalar fields to account for the electroweak symmetry breaking, can provide solutions to 
some of the questions the SM fails to answer.
Introducing additional scalar fields leads to extra Higgs-like particles, which can be either neutral or charged.
These proceedings give examples of some recent direct searches for additional Higgs bosons, using proton–proton collision data at $13$~TeV collected 
by the ATLAS experiment~\cite{atlas} in Run 2 of the LHC.

%%%%%%%%%%%%%%%%%%%%%%%%%%%%%%%%%%%%%%%%%%%%%%%%%%%%%%%%%%%%%%%%%
\section{Heavy neutral Higgs boson searches}
\begin{figure}[!h]
  \center {\includegraphics[width=0.45\textwidth]{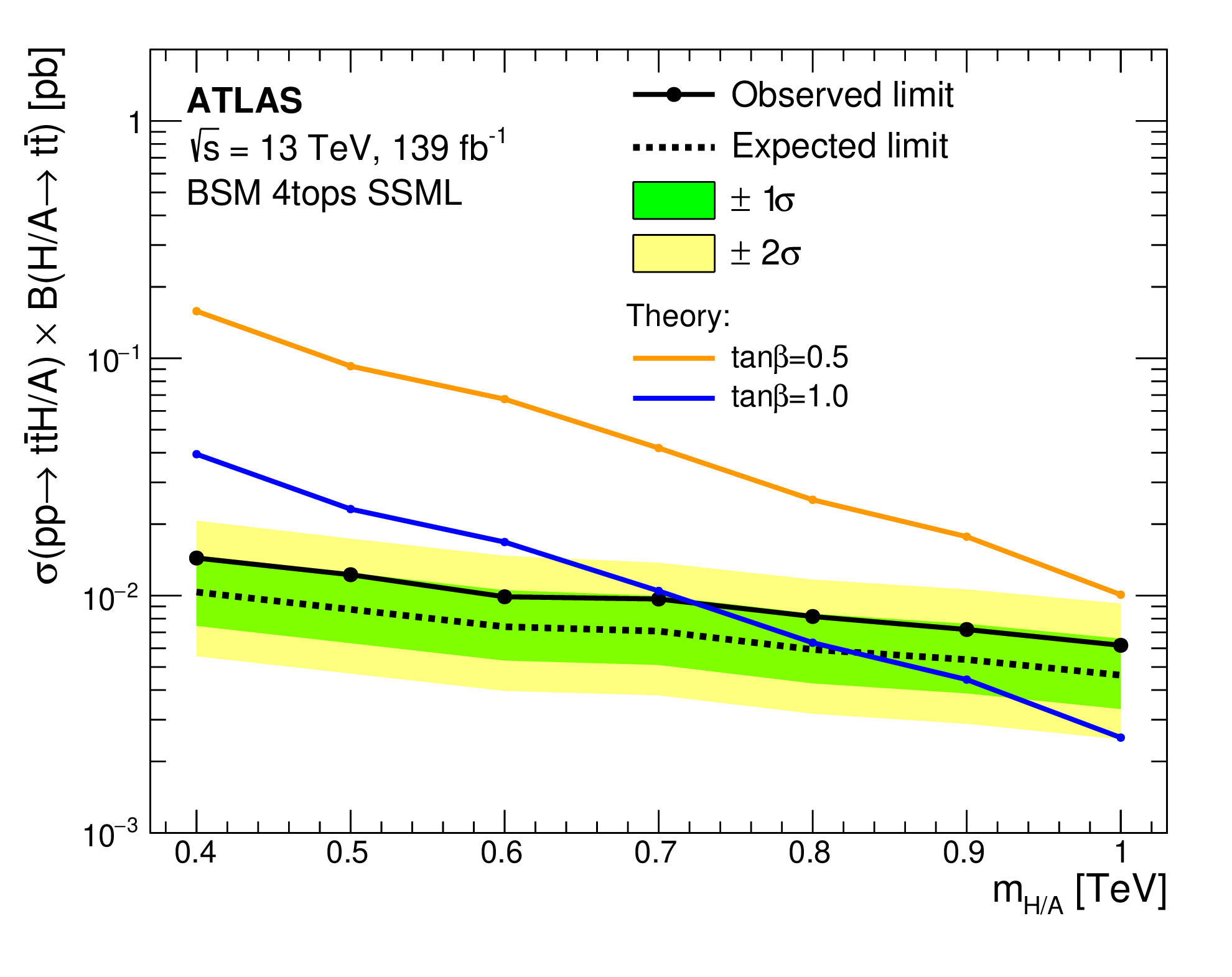}
           \includegraphics[width=0.45\textwidth]{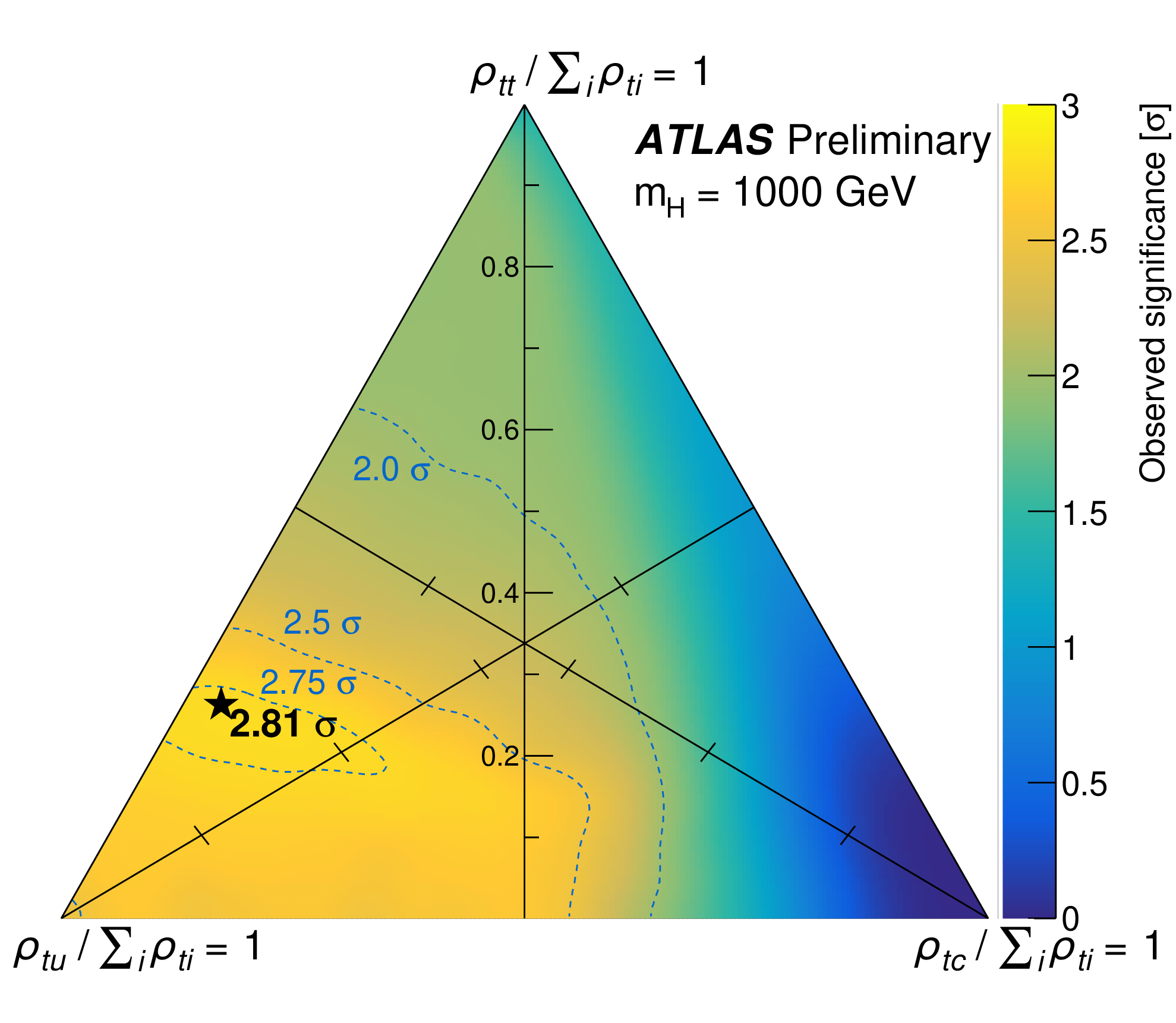}}
  \caption{ \\
   (Left) Observed (solid) and expected (dashed) 95\% CL upper limits on the $t\bar{t} H/A$ cross-section times 
   branching fraction of $H/A \rightarrow t \bar{t}$ as a function of $m_{H/A}$. 
   %The limits are estimated assuming that both a heavy scalar H boson and a pseudo-scalar A boson contribute to the $t\bar{t}t\bar{t}$
   %final state and have the same mass. 
   The green (yellow) band shows the $\pm 1\sigma$ ($\pm 2\sigma$) variation of the expected limits. 
   Theoretical predictions for two values of tan$\beta$ are shown. \cite{htt} \\
   (Right) Observed significance for $m_{H}$ = 1000 GeV as a function of the three couplings normalised to their sum.
  The star indicates the coupling configuration leading to the highest observed significance of 2.81$\sigma$. \cite{hmultil}
}
  \label{fig:htt}
\end{figure}

\textbf{\boldmath$t\bar{t}H/A \rightarrow t\bar{t} t\bar{t}$ in the multilepton final state} \newline
This search for a new heavy scalar or pseudo-scalar Higgs boson ($H/A$) produced in association with a pair of top quarks, 
with the Higgs boson decaying into a pair of top quarks ($H/A \rightarrow t \bar{t}$)~\cite{htt} is motivated by type-II 
two-Higgs-doublet models (2HDM)~\cite{2hdm}. 
%For heavy neutral Higgs bosons, the dominant decay mode is to pair of top quarks.
The $t\bar{t} H/A$  production mode provides a promising channel as inclusive searches for $H/A \rightarrow t \bar{t}$ 
are challenging due to destructive interference with the SM top pair production.
%The mass of the heavy Higgs boson 400 GeV - 1000 GeV is assumed. 
The analysis  targets a final state with exactly two leptons with same-sign electric
charges or at least three leptons.
A boosted decision trees classifier is trained to distinguish the signal from the SM background.
No significant excess of events over the SM prediction is observed and thus upper limits are placed on
the $t\bar{t} H/A$ production cross-section times the branching ratio of $H/A \rightarrow t \bar{t}$ as a function
of $m_{H/A}$, Figure~\ref{fig:htt} (Left).

\medskip
\noindent
\textbf{Heavy scalar decays in final states with multiple leptons and \boldmath$b$-jets } \newline
The presented analysis targets search for heavy scalars with flavour-violating decays in final states with multiple
leptons and $b$-jets~\cite{hmultil}.
It is motivated by general 2HDM without Z2 symmetry, where the heavy Higgs bosons feature flavour changing neutral Higgs couplings. 
Only couplings involving top quarks and two other up-type quarks ($\rho_{tt}$, $\rho_{tc}$, and $\rho_{tu}$) are considered.
The final states of interest are same-sign top quark pair, three top quarks, or four top quarks. 
A deep neural network is trained to discriminate the signal from the backgrounds.
A mild excess is observed over the SM expectation corresponding to a local significance of 2.81$\sigma$
for a signal with $m_{H}$ = 1000 GeV and $\rho_{tt}$ = 0.32, $\rho_{tc}$ = 0.05, and $\rho_{tu}$ = 0.85.
Exclusion limits at 95\% confidence are set on the mass and couplings of the heavy Higgs boson.
An observed significance for $m_{H}$ = 1000 GeV as a function of the three couplings is shown in Figure~\ref{fig:htt} (Right).
%\begin{figure}[!h]
%  \center
%  {
%  \includegraphics[width=0.45\textwidth]{hmultil.png}
%  }
%  \caption{
%  Observed significance for $m_{H}$ = 1000 GeV as a function of the three couplings normalised to their sum. 
%  The star indicates the coupling configuration leading to the highest observed significance of 2.81$\sigma$. \cite{hmultil}
%}
%  \label{fig:hmultil}
%\end{figure}

\begin{figure}[h]
  \center
  {
  \includegraphics[width=0.45\textwidth]{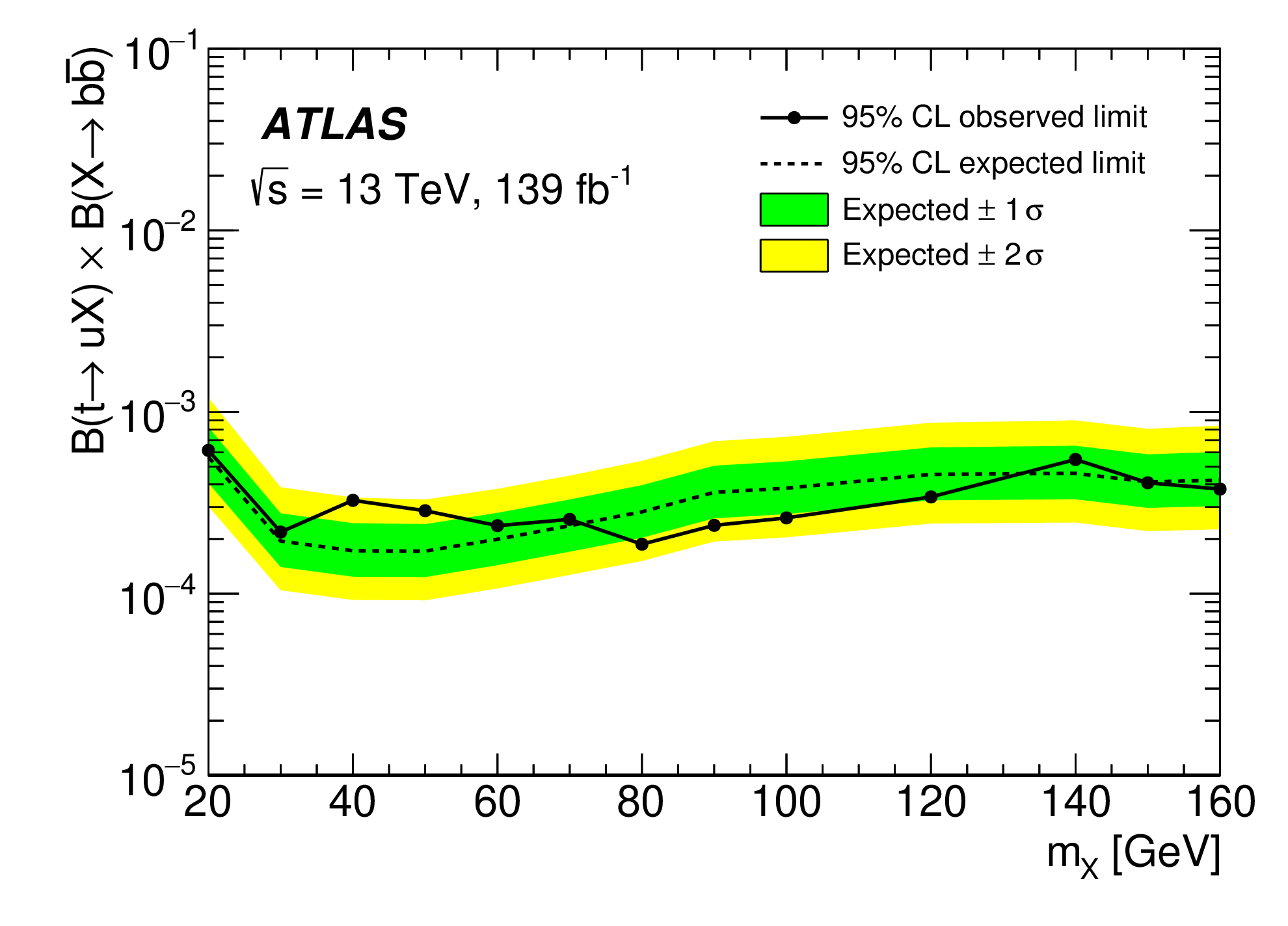}
  \includegraphics[width=0.45\textwidth]{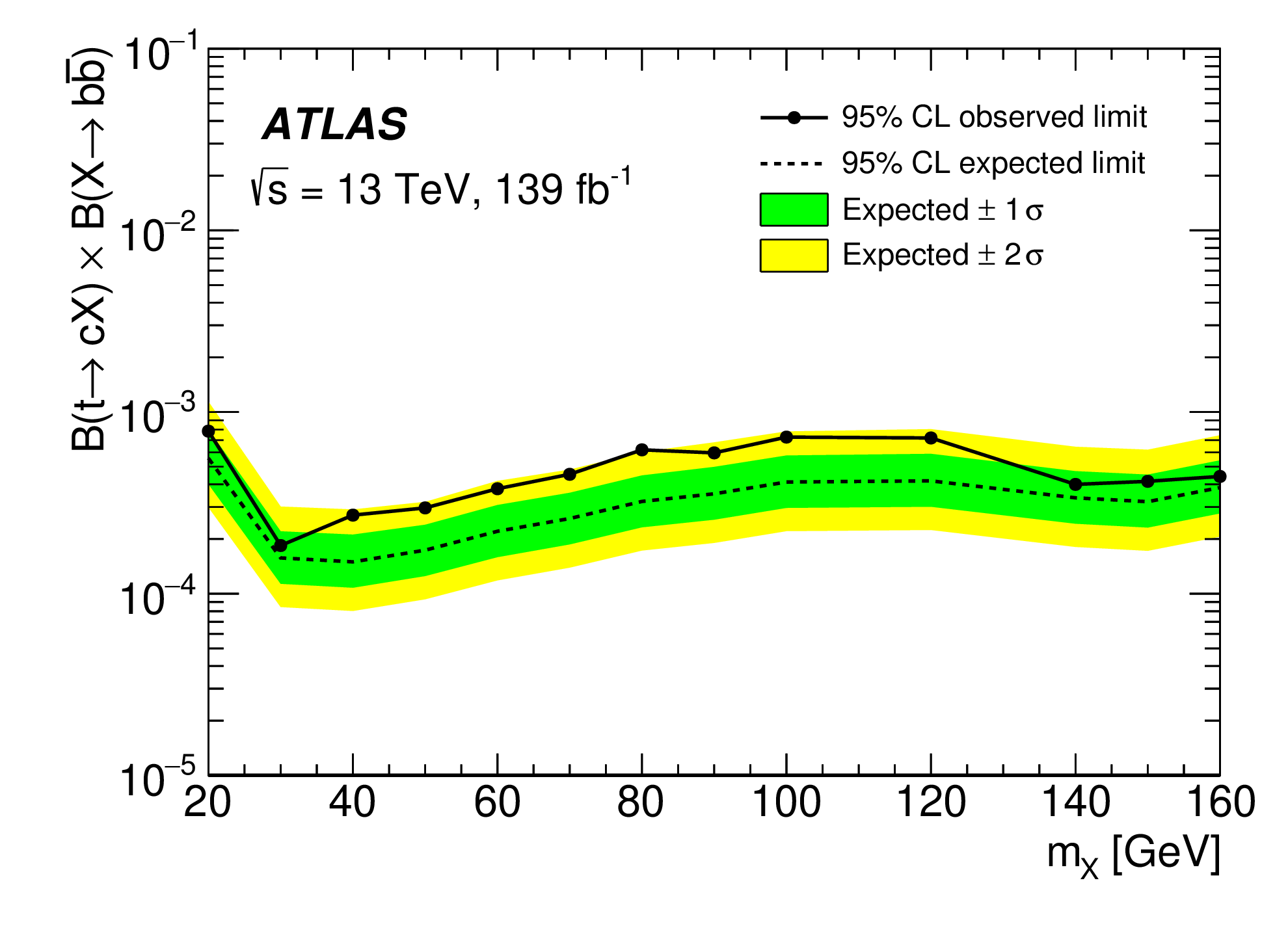}
  }
  \caption{
   Expected (dashed) and observed (solid) 95\% CL upper limits for B($t \rightarrow uX$) $\times$ B($X \rightarrow b \bar{b}$) (Left) and
   B($t \rightarrow cX$) $\times$ B($X \rightarrow b \bar{b}$) (Right). 
   The bands surrounding the expected limits show the 68\% and 95\% confidence intervals, respectively. \cite{flavon} 
}
  \label{fig:flavon}
\end{figure}

%
\noindent
\textbf{Flavour-changing neutral-current \boldmath$t \rightarrow qX$ ($q=u,c$) $\rightarrow qbb$} \newline
One of the simplest extensions to the SM is the Froggatt-Nielsen mechanism~\cite{froggatt}, which introduces a non-SM Higgs field, $X$, with flavour charge, 
the so-called flavon.
%In this model, for masses of a neutral scalar particle $X$ below 200 GeV, the leading decay mode is $X \rightarrow b \bar{b}$.
The presented analysis \cite{flavon} is a generic search for top quark pair production where one of the top quarks decays to a light scalar particle $X$,
with $X \rightarrow b \bar{b}$, and an up-type quark ($u$ or $c$).
Events are categorised according to the multiplicity of jets and $b$-jets, and a neural network is used to discriminate between signal and background processes. 
No significant excess above the expected SM background is found and the 95\% CL upper limits on B($t \rightarrow u/cX$) $\times$ B($X \rightarrow b \bar{b}$)  
are obtained as presented in Figure~\ref{fig:flavon}.
A local excess of 1.8$\sigma$ is seen in the $t \rightarrow uX$ channel at $m_X = 40$ GeV. 
Also, a roughly 2$\sigma$ excess can be seen in the $t \rightarrow cX$ observed limit over almost the entire range of $m_X$. 
This excess is not compatible with the presence of a scalar particle $X$, which would show up as a narrower, resonance-like excess.
%

%%%%%%%%%%%%%%%%%%%%%%%%%%%%%%%%%%%%%%%%%%%%%%%%%%%%%%%%%%%%%%%%%
\section{Charged Higgs boson searches}
\begin{figure}[h]
  \center {\includegraphics[width=0.45\textwidth]{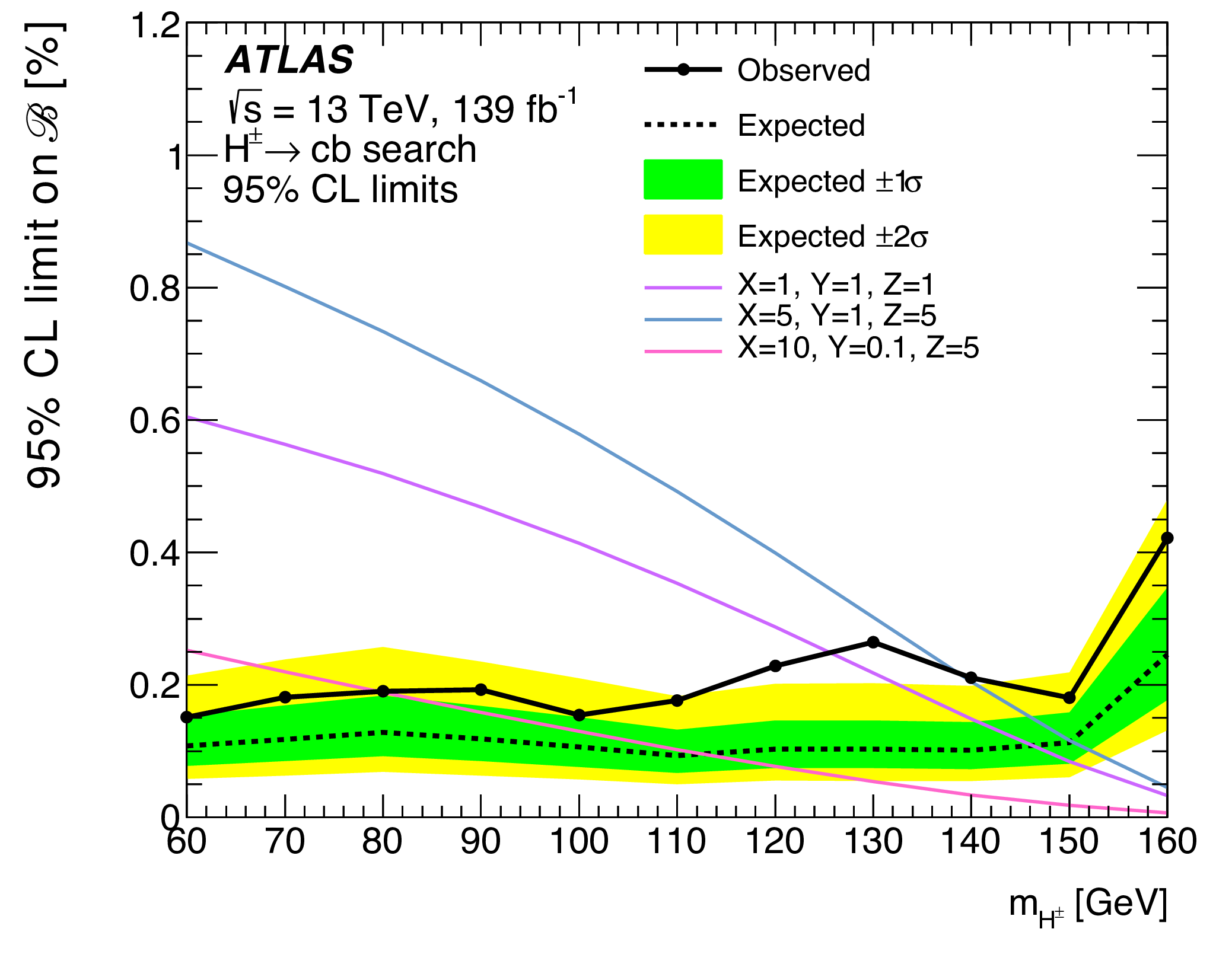}
           \includegraphics[width=0.5\textwidth]{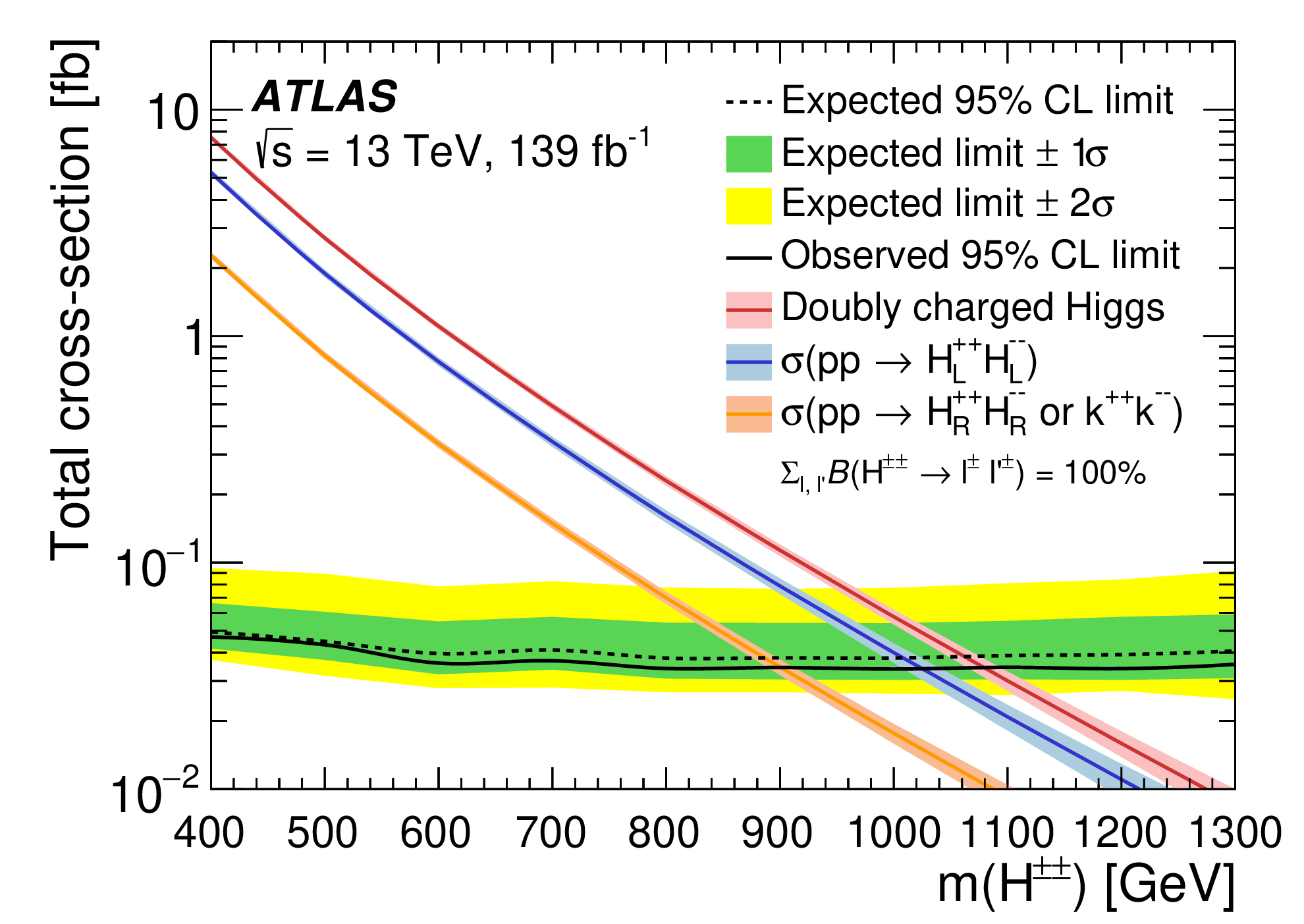}}
  \caption{ \\
   (Left) The observed (solid) and expected (dashed) 95\% CL upper limits on $B(t \rightarrow H^{\pm}b) \times B(H^{\pm} \rightarrow cb)$  as a function 
   of $m_{H^{\pm}}$. The inner green and outer yellow shaded bands show the $\pm 1\sigma$ and $\pm 2\sigma$ uncertainties of the expected limits. 
   The predictions from the 3HDM \cite{3hdm1,3hdm2} are shown, corresponding to three benchmark values for the parameters X, Y, and Z. \cite{lighthplus} \\
   (Right) Observed (solid) and expected (dashed) 95\% CL upper limits on the $H^{\pm \pm}$ pair production cross-section as a
  function of $m_{H^{\pm \pm}}$. The surrounding green and yellow bands correspond to the $\pm 1$ and $\pm2 \sigma$
  uncertainty around the combined expected limit, respectively. The theoretical signal cross-section predictions, are shown as
  blue, orange and red lines for the left-handed $H^{\pm \pm}_L$, right-handed $H^{\pm \pm}_R$, and a sum of the two LRSM chiralities,
  respectively, with the corresponding uncertainty bands. \cite{h++}
  }
  \label{fig:h+}
\end{figure}

\textbf{Light \boldmath$H^{\pm} \rightarrow cb$ produced in $t \rightarrow H^{\pm}b$ decays} \newline
The search focuses on top quark pair production, where one top quark decays into a leptonically decaying $W$ boson and 
a $b$-quark, and the other top quark may decay into a $H^{\pm}$ boson and a $b$-quark~\cite{lighthplus}. 
The $H^{\pm}$ boson decays into a $b$- and a $c$-quark are considered with $m_{H^{\pm}}=$ 60-160 GeV.
The final state consists of a single lepton, high multiplicity of jets and three or more $b$-jets. 
A mass parametrised neural network is used to separate signal from background. 
In the absence of a significant excess of data events above the SM expectation, exclusion limits at 95\% CL
on the product of branching fractions $B(t \rightarrow H^{\pm}b) \times B(H^{\pm} \rightarrow cb)$ are set in function of
$m_{H^{\pm}}$ as presented in Figure~\ref{fig:h+} (Left).
A $3\sigma$ local ($2.5\sigma$ global) broad excess is observed for $m_{H^{\pm}}=$ 130 GeV.
The excess is consistent with the $H^{\pm}$ resolution degraded by the ambiguity in choosing the correct $b$-jet
to reconstruct $H^{\pm}$ mass.

\medskip
\noindent
\textbf{\boldmath$H^{\pm \pm} \rightarrow l^{\pm} l^{\pm}$ decays} \newline 
Various beyond SM theories, for example left-right symmetric models (LRSMs)~\cite{lrsm} and the Zee–Babu neutrino mass model~\cite{zee},
predict doubly charged bosons. 
At the LHC, they would be mainly produced via Drell–Yan production. 
Presented search of $H^{\pm \pm}$ \cite{h++} focuses on small vacuum expectation value of the Higgs triplet, where only leptonic decays 
of $H^{\pm \pm}$ are relevant.
The analysis searches for same-charge lepton pairs in final states with two, three or four leptons.
The discriminant variable used in the final fit is invariant mass of the leading lepton pair.
In absence of a significant deviation from expectations, 95\% CL limits are derived.
They are presented in Figure~\ref{fig:h+} (Right).
%
%\begin{figure}[h]
%  \center{\includegraphics[width=0.6\textwidth]{fig_hpp.png}}
%  \caption{
%  Observed (solid line) and expected (dashed line) 95\% CL upper limits on the $H^{\pm \pm}$ pair production cross-section as a 
%  function of m($H^{\pm \pm}$). The surrounding green and yellow bands correspond to the $\pm 1$ and $\pm2$ standard deviation 
%  uncertainty around the combined expected limit, respectively. The theoretical signal cross-section predictions, are shown as 
%  blue, orange and red lines for the left-handed $H^{\pm \pm}_L$, right-handed $H^{\pm \pm}_R$, and a sum of the two LRSM chiralities, 
%  respectively, with the corresponding uncertainty bands. \cite{h++}
%}
%  \label{fig:h++}
%\end{figure}
%

%%%%%%%%%%%%%%%%%%%%%%%%%%%%%%%%%%%%%%%%%%%%%%%%%%%%%%%%%%%%%%%%
\section{Exotic decays of Higgs boson (125 GeV) searches}
\textbf{Dark photons from Higgs boson decays via \boldmath$ZH$ production} \newline
The dark photons are searched for in the decay of Higgs bosons $H \rightarrow \gamma \gamma_{d}$ produced through 
the $Z(\rightarrow l^{+} l^{-})H$ production mode \cite{darkpho}.
In presented analysis the vector portal is considered where the interaction results from the kinetic mixing between 
one dark and one visible Abelian gauge boson.
Both massless and light dark photons $\gamma_{d}$ (up to 40 GeV) are considered.
The final state of interest consists of two same-flavour, opposite-charge light leptons, an isolated photon and
missing transverse momentum from undetected $\gamma_{d}$.
A  boosted decision trees classifier is used to separate signal from background.
As no excess is observed with respect to the SM prediction, an observed
upper limit on the branching ratio BR($H \rightarrow \gamma \gamma_{d}$) of 2.28\% is set at
95\% CL for massless $\gamma_{d}$.

%%%%%%%%%%%%%%%%%%%%%%%%%%%%%%%%%%%%%%%%%%%%%%%%%%%%%%%%%%%%%%
\section{Concluding remarks}
Searches for additional Higgs bosons are strongly motivated by theory.
The ATLAS Collaboration has performed a wide range of such searches, covering a large variety 
of different production and decay modes.
A full review of these searches is beyond the scope of these proceedings and the reader is encouraged 
to visit the ATLAS public results webpage~\cite{atlaswebpage}.
So far, no significant hint for physics beyond the SM has been observed and also
many interesting models and regions of phase space remain unexplored.   
However there are many interesting ongoing searches besides the ones covered in this article.
The LHC Run 3 datasets will allow  to revisit some of the interesting excesses observed in the Run 2 
analyses and hopefully to discover new physics.

\bigskip
\noindent
\textbf{Funding information} This work is supported in part by the Polish Ministry of Education and Science project no. 2022/WK/08.

%%%%%%%%%%%%%%%%%%%%%%%%%%%%%%%%%%%%%%%%%%%%%%%%%%%%%%%%%%%%%%

\end{document}